\documentclass[aps,twocolumn,showpacs]{revtex4-1}

\usepackage{wallpaper}
\usepackage{dcolumn}
\usepackage{bm}
\usepackage{amsmath}
\usepackage{CJK}
\usepackage{graphics}
\usepackage{multirow}

\begin{document}

\title{Massive neutron stars and $\Lambda$-hypernuclei in relativistic mean field models}
\author{Ting-Ting~Sun$^{1}$}
\author{Cheng-Jun~Xia$^{2, 3}$}
\email{cjxia@itp.ac.cn}
\author{Shi-Sheng Zhang$^{4}$}
\author{M.~S. Smith$^{5}$}

\affiliation{$^{1}${School of Physics and Engineering and Henan Key Laboratory of Ion Beam Bioengineering, \\Zhengzhou University, Zhengzhou 450001, China}
              \\$^{2}${School of Information Science and Engineering, Ningbo Institute of Technology, Zhejiang University, Ningbo 315100, China}
              \\$^{3}${CAS Key Laboratory of Theoretical Physics, Institute of Theoretical Physics, Chinese Academy of Sciences, Beijing 100190, China}
              \\$^{4}${School of Physics and Nuclear Energy Engineering, Beihang University, Beijing 100191, China}
              \\$^{5}${Physics Division, Oak Ridge National Laboratory, Oak Ridge, Tennessee, 37831-6354, USA}}

\date{\today}

\begin{abstract}
Based on relativistic mean field (RMF) models, we study finite $\Lambda$-hypernuclei and massive neutron stars. The effective $N$-$N$ interactions PK1 and TM1 are adopted, while the $N$-$\Lambda$ interactions are constrained by reproducing the binding energy of $\Lambda$-hyperon at $1s$ orbit of $^{40}_{\Lambda}$Ca. It is found that the $\Lambda$-meson couplings follow a simple relation, indicating a fixed $\Lambda$ potential well for symmetric nuclear matter at saturation densities, i.e., around $V_{\Lambda} = -29.786$ MeV.
With those interactions, a large mass range of $\Lambda$-hypernuclei can be well described. Furthermore,  the masses of PSR J1614-2230 and PSR J0348+0432 can be attained adopting the $\Lambda$-meson couplings $g_{\sigma\Lambda}/g_{\sigma N}\gtrsim 0.73$, $g_{\omega\Lambda}/g_{\omega N}\gtrsim 0.80$ for PK1 and $g_{\sigma\Lambda}/g_{\sigma N}\gtrsim 0.81$, $g_{\omega\Lambda}/g_{\omega N}\gtrsim 0.90$ for TM1, respectively. This resolves the Hyperon Puzzle without introducing any additional degrees of freedom.
\end{abstract}


\maketitle

\section{\label{sec:intro}Introduction}
A proper equation of state (EoS) of baryonic matter is crucial to unveil the dynamics of core-collapse
supernova~\cite{Woosley2002_RMP74-1015}, neutron star properties~\cite{Lattimer2012_ARNPS62-485, Ozel2016_ApJ820-28, Ozel2016_ARAA54-401},
binary neutron star mergers~\cite{Hotokezaka2011_PRD83-124008}, and heavy-ion collisions~\cite{Danielewicz2002_Science298-1592,
Li2008_PR464-113}. Currently, the properties of nuclear matter near the saturation density $\rho_0$ are well constrained.
However, the composition of matter at higher densities is still an open question with many possibilities~\cite{Weber2005_PPNP54-193, McLerran2009_NPPS195-275}. When density is larger than 2-3 $\rho_0$, hyperons ($\Lambda$, $\Sigma$,
$\Xi$, \ldots) are created via weak reactions to lower the energy of the system. Being the lightest hyperon, $\Lambda^0$ will firstly
appear due to an attractive potential in nuclear matter~\cite{Ishizuka2008_JPG35-085201, Shen2011_ApJ197-20}, while the heavier ones can
only appear at larger densities. For neutron star matter, negatively charged hyperons such as $\Sigma^-$ and $\Xi^-$ may be important
since they can neutralize protons. However, recent studies on the quasi-free $\Sigma^-$ production
spectra~\cite{Noumi2002_PRL89-072301, Noumi2003_PRL90-049902, Saha2004_PRC70-044613} suggests that the $\Sigma^-$ potential in nuclear
matter is repulsive~\cite{Ishizuka2008_JPG35-085201}. If so, $\Sigma$-hyperons can only appear at much larger densities.

To understand the properties of baryonic matter at high densities with the possible existence of $\Lambda$-hyperons, one
needs to construct a proper $\Lambda$-baryon interaction. Many attempts were made to extract the $\Lambda$-baryon interaction, e.g., a phenomenological $N$-$\Lambda$ potential was obtained from the scattering data constraint~\cite{Bodmer1984_PRC29-684}. However, due to the limited experimental data, there are still large ambiguities for the
$N$-$\Lambda$ interaction in this scheme. Nevertheless, the structures of $\Lambda$-hypernuclei provide crucial constraints
for the $\Lambda$-baryon interaction. Based on the experimental studies on the single-$\Lambda$
hypernuclei~\cite{Hashimoto2006_PPNP57-564}, one can construct the $N$-$\Lambda$ interaction via various nuclear structure models,
e.g., shell model~\cite{Gal1971_AP63-53, Dalitz1978_AP116-167, Millener2008_NPA804-84, Millener2013_NPA914-109}, cluster
model~\cite{Motoba1983_PTP70-189, Hiyama2002_PRC66-024007, Hiyama2006_PRC74-054312, Hiyama2009_PRC80-054321, Bando1990_IJMPA05-4021},
antisymmetrized molecular dynamics~\cite{Isaka2013_PRC87-021304}, quark mean field model~\cite{Hu2014_PRC89-025802},
relativistic mean field (RMF) model~\cite{Brockmann1977_PLB69-167, Boguta1981_PLB102-93, Mares1989_ZPA333-209,Mares1994_PRC49-2472,Toki1994_PTP92-803,Song2010_IJMPE19-2538,
Tanimura2012_PRC85-014306,Wang2013_CTP60-479}, Skyrme-Hatree-Fock
model~\cite{Zhou2007_PRC76-034312}, and quark-meson coupling model~\cite{Tsushima1997_PLB411-9, Tsushima1998_NPA630-691, Guichon2008_NPA814-66}.
Note that the $\Lambda$-$\Lambda$ interaction is found to be weakly attractive in the measurement of $\Lambda$-$\Lambda$ bond energies of double-$\Lambda$ hypernuclei, such as $^6_{\Lambda\Lambda}$He~\cite{Takahashi2001_PRL87-212502}. Promising results
for the $N$-$\Lambda$ and $\Lambda$-$\Lambda$ potentials were also obtained from lattice QCD simulations~\cite{Aoki2011_PPNP66-687}.

Based on the obtained $\Lambda$-baryon interactions, great successes were achieved on understanding the properties of
$\Lambda$-hypernuclei. However, when applying those interactions to study neutron stars, the results are controversial regarding the observation of pulsars, i.e., the so called Hyperon Puzzle~\cite{Vidana2015_AIPCP1645-79}. Due to the appearance of $\Lambda$-hyperons at higher
densities, the EoS of baryonic matter becomes soft. As a consequence, the predicted mass of a neutron star can not
reach $2M_\odot$, which is not in accordance with the recently measured masses for PSR J1614-2230 ($1.928 \pm 0.017\ M_\odot$)~\cite{Demorest2010_Nature467-1081, Fonseca2016_ApJ832-167}
and PSR J0348+0432 ($2.01 \pm 0.04\ M_\odot$)~\cite{Antoniadis2013_Science340-6131}.

To solve the Hyperon Puzzle, one needs to introduce additional repulsion so that the EoS becomes stiffer
and consequently predicts neutron stars with a maximum mass exceeding $2M_\odot$. As categorized in Ref.~\cite{Vidana2016_JPCS668-012031},
there are mainly three types of mechanisms that could afford such repulsive forces: (1) the strange vector  meson $\phi$ in RMF
models which results in a repulsive interaction between hyperons~\cite{Weissenborn2012_PRC85-065802, Bednarek2012_AA543-A157,
Oertel2015_JPG42-075202, Maslov2015_PLB748-369, Maslov2016_NPA950-64}; (2) three-body forces between hyperons and nucleons in
the framework of microscopic many-body theories~\cite{Takatsuka_EPJA13-213, Vidana2011_EPL94-11002, Yamamoto2013_PRC88-022801,
Lonardoni2015_PRL114-092301, Togashi2016_PRC93-035808}; and (3) the deconfinement phase transition that happens prior to the onset density
of hyperons~\cite{Weissenborn2011_ApJ740-L14, Klahn2013_PRD88-085001, Zhao2015_PRD92-054012, Kojo2015_PRD91-045003, Masuda2016_EPJA52-1,
Li2015_PRC91-035803, Whittenbury2016_PRC93-035807, Fukushima2016_ApJ817-180}. Note that the above mentioned mechanisms involve additional
degrees of freedom which are not very well constrained. In this paper, we readjust the $\Lambda$-meson coupling strengths according to
the experimental single-$\Lambda$ binding energies of $\Lambda$-hypernuclei in the framework of the RMF model. The new interactions are
then used to calculate the mass-radius ($M$-$R$) relation of compact stars, which is compared with the observational masses
of pulsars. It is shown that the properties of $\Lambda$-hypernuclei and massive neutron stars can be well reproduced without
introducing any additional degrees of freedom.

The paper is organized as follows. In Sec.~\ref{sec:the} we present the formalism of RMF models for $\Lambda$-hypernuclei and neutron
stars. The $\Lambda$-meson coupling constants are fixed and applied to study the properties of $\Lambda$-hypernuclei and neutron
stars in Sec.~\ref{sec:num}, and a summary is given in Sec.~\ref{sec:con}.

\section{\label{sec:the}Theoretical framework}

\subsection{\label{sec:the_L}Lagrangian Density}
RMF models have been shown to be suitable for the studies of finite
(hyper)nuclei~\cite{Reinhard1989_RPP52-439, Ring1996_PPNP37_193-263,
Meng2006_PPNP57_470-563,Paar2007_RPP70-691,Meng2015_JPG42-093101,Meng2016Book,Typel1999_NPA656-331,
Vretenar1998_PRC57-R1060,Lu2011_PRC84-014328,Hagino2014_arXiv1410.7531,Sun2016_PRC94-064319} as well as baryonic matter~\cite{Glendenning2000,
Ban2004_PRC69-045805, Weber2007_PPNP59-94, Long2012_PRC85-025806,Sun2012_PRC86-014305, Wang2014_PRC90-055801, Fedoseew2015_PRC91-034307}. The starting point of the meson-exchange
RMF model for baryonic matter is the following covariant Lagrangian density
\begin{equation}
\mathcal{L}=\mathcal{L}_{N}+\mathcal{L}_{Y}+\mathcal{L}_{l}.
\label{eq:Lagrange}
\end{equation}
Here $\mathcal{L}_{N}$ is the standard RMF Lagrangian density for nucleons~\cite{Reinhard1989_RPP52-439, Ring1996_PPNP37_193-263,Meng2006_PPNP57_470-563,
Paar2007_RPP70-691} in which the couplings with the isoscalar-scalar $\sigma$, isoscalar-vector $\omega_{\mu}$, isovector-vector
$\vec{\rho}_{\mu}$ mesons, and the photon $A_{\mu}$ are included, i.e.,
\begin{eqnarray}
\mathcal{L}_{N}
 &=& \sum_{i=n,p} \bar{\psi}_{i}
       \left[  i \gamma^\mu \partial_\mu- M_i - g_{\sigma i }\sigma
              - g_{\omega i } \gamma^\mu \omega_\mu\right.\nonumber\\
 &&\mbox{}\left.- g_{\rho i } \gamma^\mu \vec{\tau}_i \cdot \vec{\rho}_\mu
              - e \gamma^\mu A_\mu \frac{1-\tau_{i,3}}{2}
              \right] \psi_{i}
\nonumber \\
 &&\mbox{}  + \frac{1}{2}\partial_\mu \sigma \partial^\mu \sigma - \frac{1}{2}m_\sigma^2 \sigma^2-\frac{1}{3}g_2 \sigma^3 -\frac{1}{4}g_3 \sigma^4
\nonumber \\
 &&\mbox{} - \frac{1}{4} \Omega_{\mu\nu}\Omega^{\mu\nu} + \frac{1}{2}m_\omega^2 \omega_\mu\omega^\mu +
 \frac{1}{4}c_3 \left(\omega_\mu\omega^\mu\right)^2
\nonumber \\
 &&\mbox{} - \frac{1}{4} \vec{R}_{\mu\nu}\cdot\vec{R}^{\mu\nu}
     + \frac{1}{2}m_\rho^2 \vec{\rho}_\mu\cdot\vec{\rho}^\mu
 - \frac{1}{4} F_{\mu\nu}F^{\mu\nu},
     \label{eq:Lagrange_NN}
\end{eqnarray}
where $M_i~(i=n,p)$ denotes the nucleon mass, $\vec{\tau}_i$ is the isospin with its 3rd component $\tau_{i,3}$, and
$m_{\sigma}(g_{\sigma i})$, $m_{\omega}(g_{\omega i})$ and $m_{\rho}(g_{\rho i})$
are the masses (coupling constants) for the $\sigma$-, $\omega$-, and $\rho$-mesons, respectively. Note that $g_{2}$, $g_{3}$, and $c_3$ are parameters introduced in the nonlinear self-coupling terms.
The field tensors of the vector mesons $\Omega$ and $\vec{R}$ and photons $F$ are defined as
\begin{subequations}
\begin{eqnarray}
\Omega_{\mu\nu} &=& \partial_\mu \omega_\nu - \partial_\nu \omega_\mu, \\
\vec{R}_{\mu\nu}
  &=& \partial_\mu \vec{\rho}_\nu - \partial_\nu \vec{\rho}_\mu, \\
F_{\mu\nu} &=& \partial_\mu A_\nu - \partial_\nu A_\mu.
\end{eqnarray}
\end{subequations}
We adopt the arrows to indicate vectors in isospin space.

The Lagrangian density $\mathcal{L}_{Y}$ represents the contributions from hyperons~\cite{Wang2013_CTP60-479,Lu2011_PRC84-014328,Hagino2014_arXiv1410.7531}. Since $\Lambda$ hyperons are charge neutral with isospin $\vec{\tau}=0$, only the couplings with $\sigma$- and $\omega$-mesons are included, i.e.,
\begin{eqnarray}
\mathcal{L}_{Y}&=&\bar{\psi}_{\Lambda}
       \left[i \gamma^\mu \partial_\mu- M_\Lambda - g_{\sigma \Lambda}\sigma
              - g_{\omega \Lambda} \gamma^\mu \omega_\mu\right.\nonumber\\
              &&\left.-\frac{f_{\omega\Lambda}}{2M_{\Lambda}}\sigma^{\mu\nu}\partial_{\nu}\omega_{\mu}\right]\psi_{\Lambda},
              \label{eq:Lagrange_LN}
\end{eqnarray}
where $M_{\Lambda}$ is the mass of $\Lambda$ hyperon, $g_{\sigma\Lambda}$ and $g_{\omega\Lambda}$ are coupling
constants with $\sigma$- and $\omega$-mesons, respectively. The last term in $\mathcal{L}_{Y}$ is the tensor coupling with $\omega$-meson, which is related with the s.p.~spin-orbit splitting.

Meanwhile, the Lagrangian density $\mathcal{L}_{l}$ is for $e$ and $\mu$ leptons with
\begin{equation}
\mathcal{L}_{l} =\sum_{i=e,\mu} \bar{\psi}_i \left[ i \gamma^\mu \partial_\mu + e \gamma^\mu A_\mu - M_i\right]\psi_i,
\end{equation}
and $M_{i}(i=e,\mu)$ are their masses.

For a system with time-reversal symmetry, the space-like components of the vector fields $\omega_\mu$ and $\vec{\rho}_\mu$ vanish,
leaving only the time components $\omega_0$ and $\vec{\rho}_0$. Meanwhile, the charge conservation guarantees that only the 3rd
component $\rho_{0,3}$ in the isospin space of $\vec{\rho}_0$ survives. Adopting the mean field and no-sea approximations, the single particle
(s.p.) Dirac equations for baryons and the Klein-Gordon equations for mesons and photon can be obtained by a variational procedure.

\subsection{\label{sec:the_HN} Finite $\Lambda$-hypernuclei}
To investigate finite $\Lambda$-hypernuclei, we neglect leptons since their contributions are comparatively small.
In the spherical cases, the Dirac spinor for baryons can be expanded as
\begin{equation}
 \psi_{n\kappa m}({\bm r}) =\frac{1}{r}
 \left(\begin{array}{c}
   iG_{n\kappa}(r) \\
    F_{n\kappa}(r) {\bm\sigma}\cdot{\hat{\bm r}} \\
 \end{array}\right) Y_{jm}^l(\theta,\phi)\:,
\label{EQ:RWF}
\end{equation}
with $G_{n\kappa}(r)/r$ and $F_{n\kappa}(r)/r$ being the radial wave functions for the upper and lower components, while $Y_{jm}^l(\theta,\phi)$
is the spinor spherical harmonics. The quantum number $\kappa$ is defined by the angular momenta $(l,j)$ as $\kappa=(-1)^{j+l+1/2}(j+1/2)$.

Based on the variational method, the Dirac equation for the radial wave functions of baryons ($i=n,p,\Lambda$) is obtained as
\begin{equation}
 \left(\begin{array}{cc}
  V_i+S_i                             & {\displaystyle -\frac{d}{dr}+\frac{\kappa}{r}+T_i}\\
  {\displaystyle \frac{d}{dr}+\frac{\kappa}{r}+ T_i} & V_i-S_i-2M_i                       \\
 \end{array}\right)
 \left(\begin{array}{c}
  G_{n\kappa} \\
  F_{n\kappa} \\
 \end{array}\right)
 = \varepsilon_{n\kappa}
 \left(\begin{array}{c}
  G_{n\kappa} \\
  F_{n\kappa} \\
 \end{array}\right) \:,
\label{EQ:RDirac}
\end{equation}
with the s.p.~energy $\varepsilon_{n\kappa}$ and the mean field scalar, vector and tensor potentials
\begin{subequations}
\begin{eqnarray}
 S_i&=&g_{\sigma i}\sigma,\\
 V_i&=& g_{\omega i}\omega_0+g_{\rho i}\tau_{i,3}\rho_{0,3}+\frac{1}{2}e(1-\tau_{i,3})A_0,\\
 T_i&=&-{\displaystyle \frac{f_{\omega i}}{2M_i}}\partial_r\omega_0.
\end{eqnarray}%
 \label{EQ:potential}%
\end{subequations}%
Note that the terms related with $\rho_{0,3}$ and $A_0$ in Eq.~(\ref{EQ:potential}b) are zero
for $\Lambda$ hyperons, while the tensor potential in Eq.~(\ref{EQ:potential}c) is zero for nucleons.

The Klein-Gordon equations for mesons and photons are,
\begin{equation}
(\partial^\mu\partial_\mu+m_{\phi}^2)\phi=S_{\phi},
\label{eq:K-G}
\end{equation}
with source terms
\begin{equation}
S_{\phi}=
\left\{
  \begin{array}{ll}
    \sum \limits_{i=n,p,\Lambda}-g_{\sigma i}\rho_{s i}-g_{2}\sigma^2-g_3\sigma^3, & \hbox{$\phi=\sigma$;} \\
    \sum \limits_{i=n,p,\Lambda} g_{\omega i}\rho_{v i}
    +{\displaystyle \frac{f_{\omega\Lambda}}{2M_{\Lambda}}}\partial_kj_{T\Lambda}^{0k}-c_3\omega_0^3, & \hbox{$\phi=\omega$;} \\
    \sum \limits_{i=n,p}g_{\rho i}\tau_{i,3}\rho_{v i}, & \hbox{$\phi=\rho$;} \\
    e\rho_c, & \hbox{$\phi=A$.}
  \end{array}
\right.
\label{EQ:source}
\end{equation}
where $\rho_{si}$ and $\rho_{vi}$ are the scalar and baryon densities for nucleons, $\rho_c$ is the charge density for protons, and $j_{T\Lambda}^{0k}$ is the tensor density for $\Lambda$ hyperons.

With the radial wave functions,
the densities for the baryons in Eq.~(\ref{EQ:source}) can be expressed as
\begin{subequations}
\begin{eqnarray}
 \rho_{s i}(r) &=& \frac{1}{4\pi r^2}\sum_{k=1}^{A_i}
 \left[|G_{k i}(r)|^2-|F_{k i}(r)|^2\right] \:,
\\
 \rho_{v i}(r) &=& \frac{1}{4\pi r^2}\sum_{k=1}^{A_i}
 \left[|G_{k i}(r)|^2+|F_{k i}(r)|^2\right] \:,%
\\
 \rho_{c}(r) &=& \frac{1}{4\pi r^2}\sum_{k=1}^{A_p}
 \left[|G_{k p}(r)|^2+|F_{k p}(r)|^2\right] \:,%
\\
{\bm j}_{T\Lambda}^{0}&=&\frac{1}{4\pi r^2}\sum_{k=1}^{A_{\Lambda}}
 \left[2G_{k \Lambda}(r)F_{k\Lambda}(r)\right]{\bm n}\:,%
\end{eqnarray}%
\label{EQ:Density}%
\end{subequations}%
where ${\bm n}$ is the angular unit vector. The baryon number $A_i (i=n,p,\Lambda)$ can be calculated by integrating the baryon density $\rho_{v i}(r)$ in coordinate space as
\begin{equation}
 A_i = \int 4\pi r^2\mbox{d}r\; \rho_{v i}(r) \:.
\label{e:axi}
\end{equation}

For given $N$-$N$ and $N$-$\Lambda$ effective interactions, we solved the Dirac Eq.~(\ref{EQ:RDirac}), mean field potentials Eq.~(\ref{EQ:potential}), Klein-Gordon Eq.~(\ref{eq:K-G}),
and densities Eq.~(\ref{EQ:Density}) in the RMF model by iteration in coordinate space with a box size of $R=20~{\rm fm}$ and a step size of $0.05~{\rm fm}$.

\subsection{\label{sec:the_nstar}Neutron Stars}
To investigate neutron stars with RMF models, the procedure is similar to that used for finite nuclei. However, a neutron star is comprised
of approximately $10^{57}$ baryons with the radius being typically around $10~$km, which is much larger than a nucleus. When we
consider a large number of particles in a large system, the exact solutions of the boundary problem can be replaced by plane waves.
Then the meson fields in neutron star matter are determined by
\begin{subequations}
\begin{eqnarray}
m_\sigma^2 \sigma &=& \sum_{i=n,p,\Lambda} -g_{\sigma i} \rho_{s i} - g_2 \sigma^2 - g_3 \sigma^3, \\
m_\omega^2 \omega_0 &=& \sum_{i=n,p,\Lambda} g_{\omega i} \rho_{v i} - c_3 \omega_0^3, \\
m_\rho^2 \rho_{0,3} &=& \sum_{i=n,p} g_{\rho i}\tau_{i,3} \rho_{v i}, 
\end{eqnarray}
\end{subequations}
with the source currents of baryon $i$
\begin{subequations}
\begin{eqnarray}
\rho_{v i} &=& \langle \bar{\psi}_i\gamma^0 \psi_i \rangle = \frac{g_i\nu_i^3}{6\pi^2}, \label{eq:ni} \\
\rho_{s i} &=&\langle \bar{\psi}_i\psi_i \rangle
=\frac{g_i m_i^{*3}}{4\pi^2} \left[x_i\sqrt{x_i^2+1} - \mathrm{arcsh}(x_i)\right].~~~~~~~ \label{eq:ns}
\end{eqnarray}
\end{subequations}
Here we have defined $x_i\equiv \nu_i/m_i^*$ with $\nu_i$ being the Fermi momentum and $g_i=2$ the degeneracy factor for particle type $i$.
The effective masses of baryons are given as $m_i^*\equiv m_i + g_{\sigma i} \sigma$, while the meson fields $\sigma$, $\omega_0$, and $\rho_{0,3}$
are obtained as their mean values. Note that the charge density is zero in neutron star matter since the local charge neutrality condition should be fulfilled, i.e.,
\begin{equation}
  \sum_i q_i \rho_{v i} = 0, \label{eq:Chntr}
\end{equation}
with $q_i$ being the charge of particle type $i=(p,e^-,\mu^-)$. Also, the tensor potential $T_{\Lambda}$ and density ${\bm j}_{T\Lambda}^{0k}$ for $\Lambda$ hyperons vanish in uniform neutron stars.

At zero temperature, the energy density can be determined by
\begin{eqnarray}
E &=& \sum_{i}\varepsilon_i(\nu_i, m_i^*) +
    \frac{1}{2}m_\sigma^2 \sigma^2 + \frac{1}{3}g_2 \sigma^3 + \frac{1}{4}g_3 \sigma^4
\nonumber \\
&&\mbox{}
  + \frac{1}{2}m_\omega^2 \omega_0^2 + \frac{3}{4}c_3 \omega_0^4
  + \frac{1}{2}m_\rho^2 \rho_{0,3}^2,
\label{eq:E}
\end{eqnarray}
with the kinetic energy density of fermion $i$ as
\begin{eqnarray}
\varepsilon_i(\nu_i, m_i^*) &=& \int_0^{\nu_i} \frac{g_i p^2}{2\pi^2} \sqrt{p^2+m_i^{*2}}\mbox{d}p \label{eq:ei0}\\
&=&  \frac{g_i m_i^{*4}}{16\pi^{2}} \left[x_i(2x_i^2+1)\sqrt{x_i^2+1}-\mathrm{arcsh}(x_i) \right].
\nonumber
\end{eqnarray}
Note that effective masses of leptons are their own, i.e., $m_{e,\mu}^*= m_{e,\mu}$. The chemical potentials for baryons $\mu_b (b=n,p,\Lambda)$
and leptons $\mu_l (l=e,\mu)$ are obtained from
\begin{subequations}
\begin{eqnarray}
\mu_b&=& g_{\omega b} \omega_0
              + g_{\rho b}\tau_{b,3} \rho_{0,3}
              + \sqrt{\nu_b^2+{m_b^*}^2},
\label{eq:chem_B} \\
\mu_l &=&  \sqrt{\nu_l^2+m_l^2}.
\label{eq:chem_l}
\end{eqnarray}
\end{subequations}
Then the pressure is determined by
\begin{equation}
P = \sum_{i} \mu_i \rho_{v i} - E. \label{eq:pressure}
\end{equation}

To reach the lowest energy, particles will undergo weak reactions until
the $\beta$-equilibrium condition is satisfied, i.e.,
\begin{equation}
\mu_\Lambda= \mu_n,~~ \mu_e= \mu_n- \mu_p,~~ \mu_\mu = \mu_e.  \label{eq:weakequi}
\end{equation}
The EoS of neutron star matter is obtained with Eqs.~(\ref{eq:E}) and (\ref{eq:pressure}), then the structure of a neutron star is
determined by solving the Tolman-Oppenheimer-Volkov (TOV) equation
\begin{equation}
\frac{\mbox{d}P}{\mbox{d}r}
=-\frac{G M E}{r^2}
  \frac{(1+P/E)(1+4\pi r^3 P/M)} {1-2G M/r},  \label{eq:TOV}
\end{equation}
with the subsidiary condition
\begin{equation}
M(r) = \int_0^r 4\pi E r^2 \mbox{d}r. \label{eq:m_star}
\end{equation}
Here the gravitational constant $G=6.707\times 10^{-45}\ \mathrm{MeV}^{-2}$.

\section{\label{sec:num}Results and discussions}

\begin{figure}
\includegraphics[width=\linewidth]{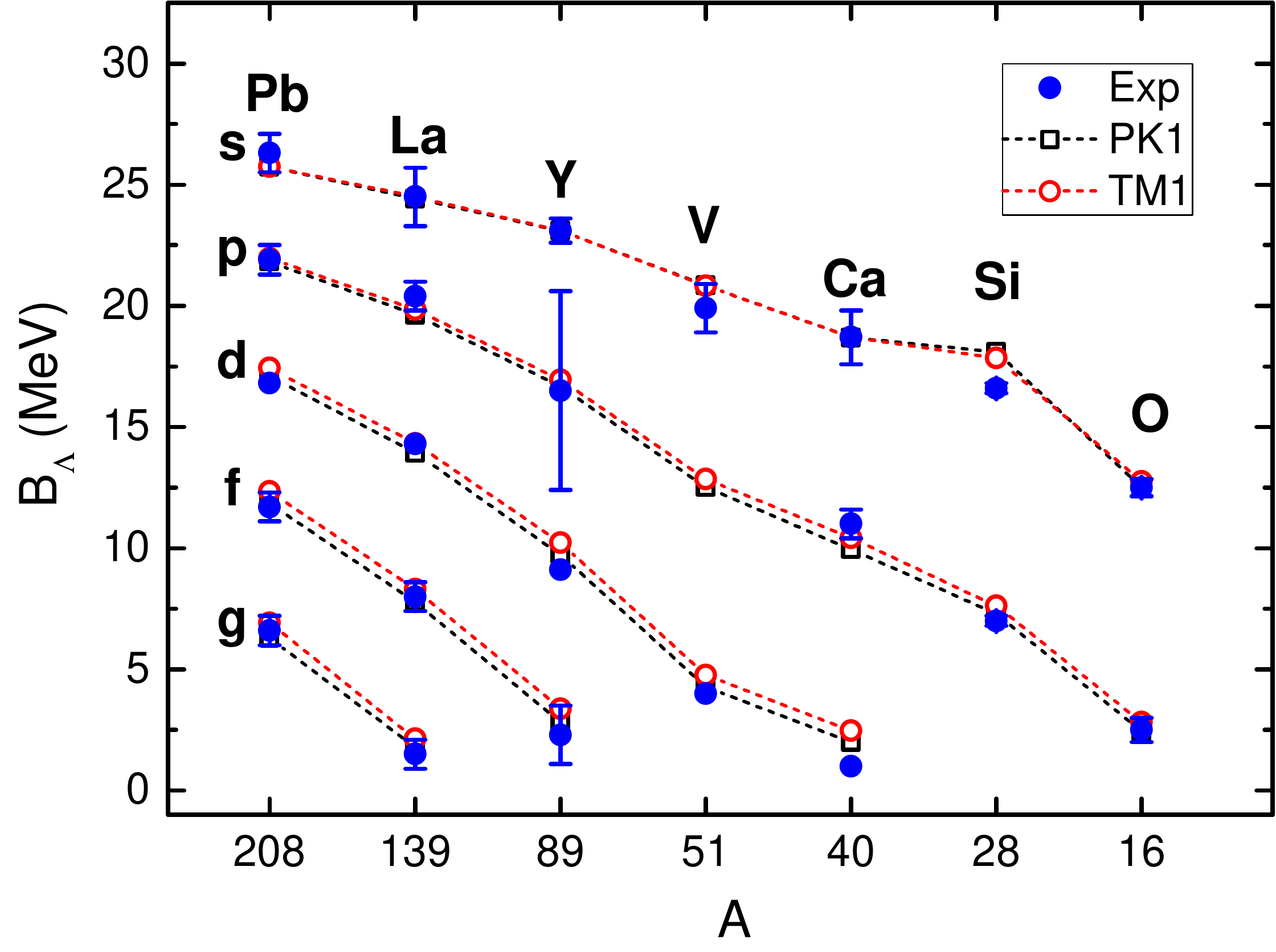}
\caption{\label{Fig:B_Lam} (Color online) The predicted single-$\Lambda$ binding energies of $\Lambda$-hypernuclei based on the effective
interactions PK1~\cite{Long2004_PRC69-034319} and TM1~\cite{Sugahara1994_NPA579-557}, which are compared with the experimental
data~\cite{Hashimoto2006_PPNP57-564}.}
\end{figure}

We carried out extensive calculations to investigate the properties of $\Lambda$-hypernuclei and neutron stars based on the effective $N$-$N$
interactions PK1~\cite{Long2004_PRC69-034319} and TM1~\cite{Sugahara1994_NPA579-557}. Note that the effective interaction PK1 has been
widely adopted in our previous studies on the properties of ordinary nuclei and hypernuclei, while the effective interaction TM1 has been commonly
used for supernova simulations, i.e., the Shen EoSs~\cite{Shen2011_ApJ197-20}.
For the $N$-$\Lambda$ interactions, the scalar coupling constant $\alpha_{\sigma\Lambda} \equiv g_{\sigma\Lambda}/g_{\sigma N}$ is constrained by reproducing the experimental binding energies of $\Lambda$ hyperon in the $1s_{1/2}$ state of hypernucleus $^{40}_{\Lambda}$Ca~($B_{\Lambda}=18.7~$MeV)~\cite{Hashimoto2006_PPNP57-564}, which are $0.618$ and $0.620$ for the effective interaction PK1 and TM1, respectively; the vector coupling constant $\alpha_{\omega\Lambda}\equiv g_{\omega\Lambda}/g_{\omega N} = 0.666$ is fixed according to the naive quark model~\cite{Dover1984_PPNP12-171}; and the tensor coupling constant $f_{\omega\Lambda}=-1.0 g_{\omega\Lambda}$ is taken as in Ref.~\cite{Mares1994_PRC49-2472}. The masses of $\Lambda^0$, $e^-$ and $\mu^-$ are taken from the Particle Data Group~\cite{Olive2014_CPC38-090001}.
With those $N$-$N$ and $N$-$\Lambda$ interactions, we calculated the mass dependence of the single-$\Lambda$ binding energies of $\Lambda$-hypernuclei and present the results in Fig.~\ref{Fig:B_Lam} along with their experimental values. Note that the theoretical single-$\Lambda$ binding energies in Fig.~\ref{Fig:B_Lam} are the average values of those for spin up and spin down orbits. It is found that, with the present $N$-$\Lambda$ interactions based on the PK1 and TM1 effective $N$-$N$ interactions, the RMF model can well describe the hypernuclei in a large mass range of $A=16\sim 208$, especially for the heavy hypernucleus $^{208}_\Lambda$Pb.

\begin{figure}
\includegraphics[width=\linewidth]{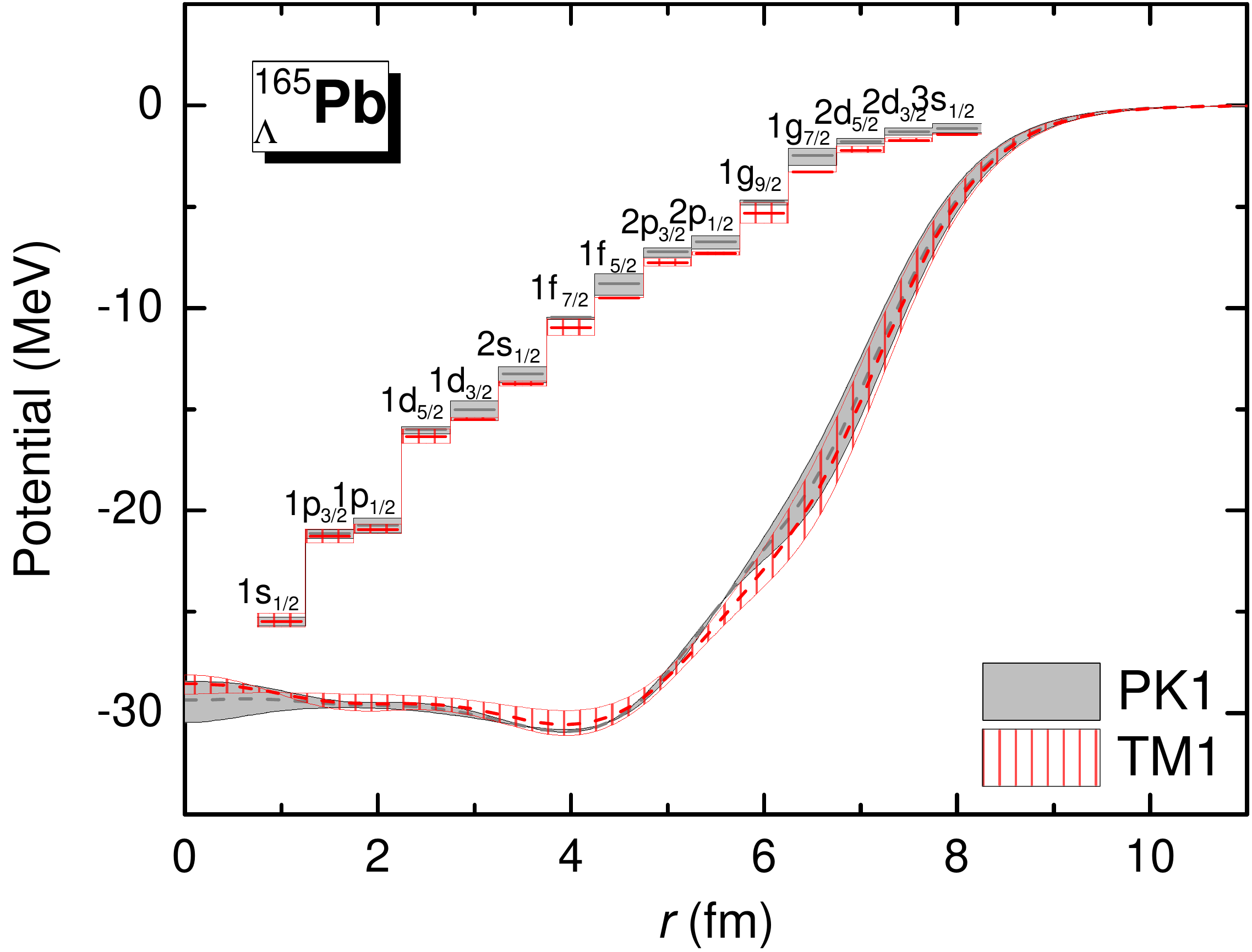}
\caption{\label{Fig:Potentail_Pb} (Color online) The mean field potential and single-particle levels of $\Lambda$ in $^{165}_\Lambda$Pb.}
\end{figure}

\begin{table}
\caption{\label{table:coupling} The coupling constants $\alpha_{\omega\Lambda}$ and $\alpha_{\sigma\Lambda}$ for $N$-$\Lambda$ interactions. The corresponding onset densities $\rho_0^\Lambda$
(in fm${}^{-3}$) and chemical potentials $\mu_0^\Lambda$ (in MeV) for $\Lambda$ hyperons are listed here as well.}
\begin{tabular}{c|ccc|ccc} \hline \hline
  &   \multicolumn{3}{c|}{PK1} & \multicolumn{3}{c}{TM1}  \\ \hline
$\alpha_{\omega\Lambda}$  & $\alpha_{\sigma\Lambda}$ & $\rho_0^\Lambda$ & $\mu_0^\Lambda$ & $\alpha_{\sigma\Lambda}$  & $\rho_0^\Lambda$ & $\mu_0^\Lambda$ \\\hline
0.60  &  0.565  &  0.296 &  1105.03 &  0.567  &  0.308 &  1104.96 \\
0.65  &  0.605  &  0.300 &  1109.19 &  0.607  &  0.312 &  1109.22 \\
0.70  &  0.645  &  0.304 &  1113.75 &  0.647  &  0.317 &  1113.76 \\
0.75  &  0.686  &  0.309 &  1118.67 &  0.687  &  0.323 &  1118.74 \\
0.80  &  0.726  &  0.313 &  1124.05 &  0.727  &  0.328 &  1124.15 \\
0.85  &  0.767  &  0.319 &  1129.96 &  0.767  &  0.334 &  1130.03 \\
0.90  &  0.807  &  0.325 &  1136.50 &  0.807  &  0.341 &  1136.49 \\
0.95  &  0.847  &  0.331 &  1143.73 &  0.847  &  0.348 &  1143.67 \\
1.00  &  0.888  &  0.338 &  1151.97 &  0.887  &  0.357 &  1151.62 \\
\hline
\end{tabular}
\end{table}

However, the aforementioned parameter sets are not unique. In fact, as long as the depth of $\Lambda$ mean field potential
\begin{equation}
  V_\Lambda\equiv g_{\sigma\Lambda}\sigma + g_{\omega\Lambda}\omega_0  \label{eq:Vlam}
\end{equation}
is fixed, the predicted single-$\Lambda$ binding energies barely vary with $\alpha_{\sigma\Lambda}$ or $\alpha_{\omega\Lambda}$~\cite{Wang2013_CTP60-479}. This is shown clearly in Fig.~\ref{Fig:Potentail_Pb}, where the mean field potential and
single-particle levels of $\Lambda$ hyperon in $^{165}_\Lambda$Pb are presented at various choices of $\alpha_{\sigma\Lambda}$ and $\alpha_{\omega\Lambda}$.
Specifically, the shaded region in Fig.~\ref{Fig:Potentail_Pb} are obtained by varying $\alpha_{\omega\Lambda}$ from 0.666
to 1, while the dashed lines represent the central values obtained with $\alpha_{\omega\Lambda} = 0.85$.
The $\sigma$-$\Lambda$ couplings are fixed to give $V_\Lambda= -29.786$ MeV for symmetric nuclear matter at saturation densities.
For various choices of $\alpha_{\omega\Lambda}$, the corresponding values of $\alpha_{\sigma\Lambda}$ that reproduce the binding energies
of $\Lambda$-hypernuclei are listed in Table~\ref{table:coupling}. It is found
that the $\Lambda$ potential varies little for different $\Lambda$-meson couplings or nuclear effective interactions. Correspondingly,
the single-$\Lambda$ binding energies are well constrained within 1 MeV for $^{165}_\Lambda$Pb as indicated in Fig.~\ref{Fig:Potentail_Pb}.
Similar behaviors are observed for other $\Lambda$-hypernuclei as well.

\begin{figure}
\includegraphics[width=\linewidth]{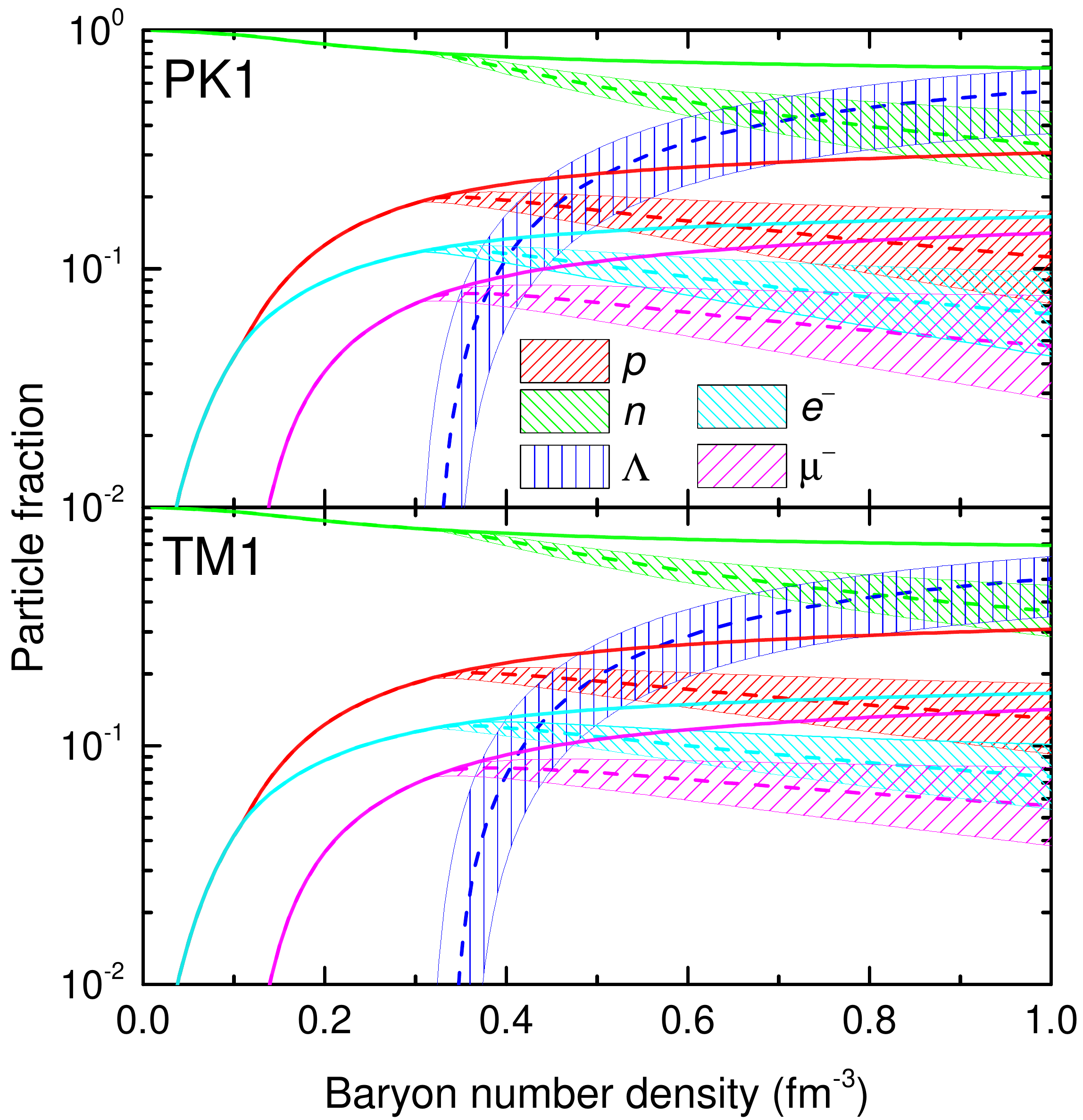}
\caption{\label{Fig:ni} (Color online) Particle fractions $n_i/n$ of baryons ($p, n, \Lambda$) and leptons ($e, \mu$) in
neutron star matter. The shaded region are obtained by varying $\alpha_{\omega\Lambda}$ from 0.666 to 1, while the dashed lines in the shadows and
solid lines above the shadows correspond to the results obtained with $\alpha_{\omega\Lambda} = 0.85$ and without hyperons, respectively; the same convention is
adopted for the following figures.}
\end{figure}

At a given total baryon number density $n$, the properties of neutron star matter can be obtained by simultaneously fulfilling the conditions
of baryon number conservation with $n=\rho_{v n}+\rho_{v p}+\rho_{v \Lambda}$, charge neutrality in Eq.~(\ref{eq:Chntr}), and
chemical equilibrium in Eq.~(\ref{eq:weakequi}). As was done for $^{165}_\Lambda$Pb in Fig.~\ref{Fig:Potentail_Pb}, we do not specify the exact
values for $\alpha_{\sigma\Lambda}$ and $\alpha_{\omega\Lambda}$, but rather varying $\alpha_{\omega\Lambda}$ from 0.666 to 1 while
$\alpha_{\sigma\Lambda}$ is determined by Eq.~(\ref{eq:Vlam}) with $V_\Lambda= -29.786$ MeV. The particle number density for each species
is determined by Eq.~(\ref{eq:ni}), where the particle fractions are presented in Fig.~\ref{Fig:ni} as functions of the total baryon number
density $n$. As indicated in Table~\ref{table:coupling}, varying $\alpha_{\omega\Lambda}$ has minor impact on hyperonic matter
at lower densities, where the onset density for $\Lambda$ increases from $0.30$ to $0.36\ \mathrm{fm}^{-3}$ as we increase the values of
$\alpha_{\omega\Lambda}$. At larger densities, the neutron star matter is dominated by $\Lambda$ starting at $n = 0.6$ to $1.4\ \mathrm{fm}^{-3}$,
depending on the adopted interactions.

\begin{figure}
\includegraphics[width=\linewidth]{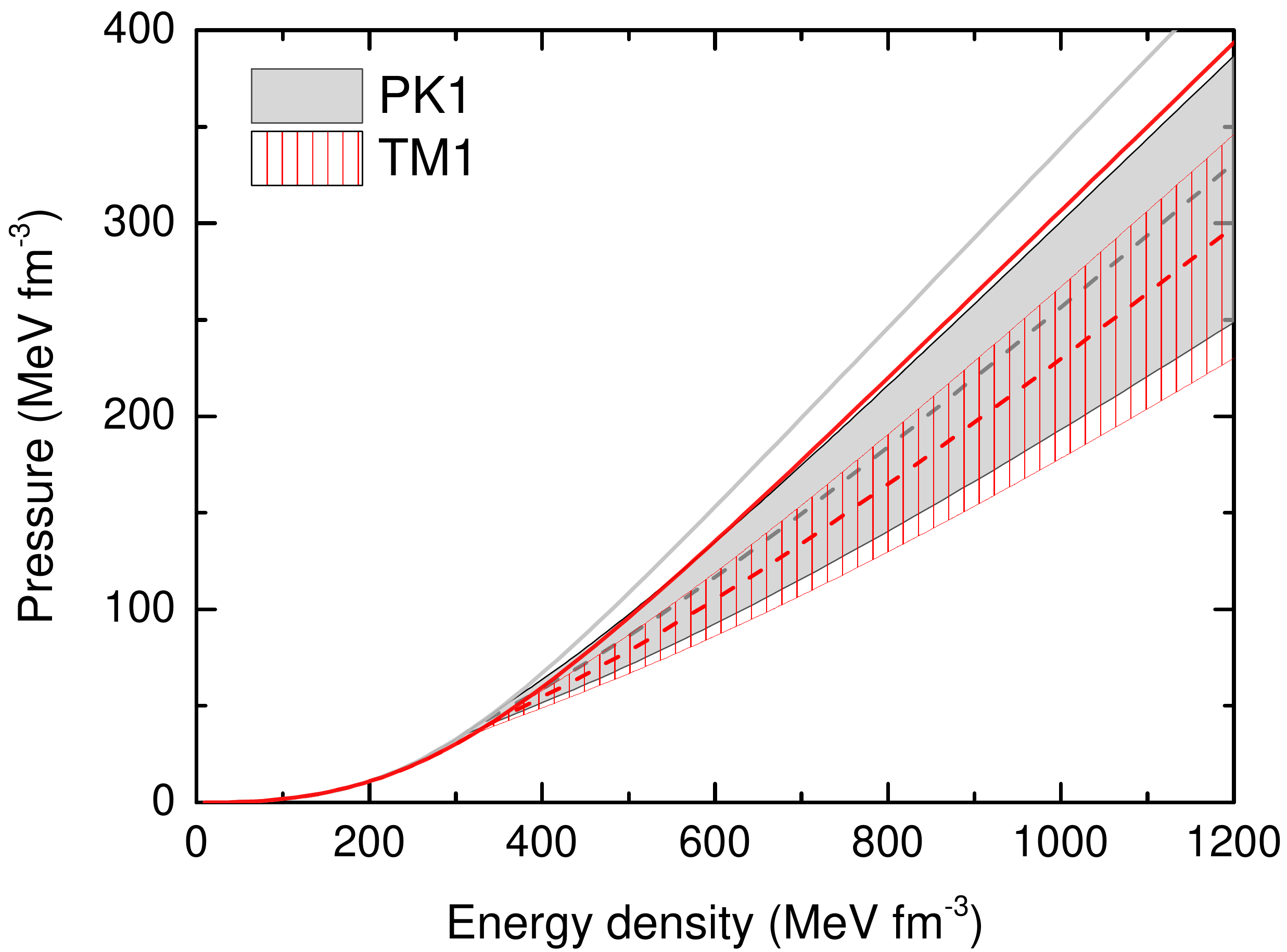}
\caption{\label{Fig:EoS} (Color online) The pressure $P$ of neutron star matter as functions of the energy density $E$.}
\end{figure}

The energy density $E$ and pressure $P$ of neutron star matter are obtained with Eqs.~(\ref{eq:E}) and (\ref{eq:pressure}) along with the
subsidiary equations in Sec.~\ref{sec:the_nstar}. In Fig.~\ref{Fig:EoS}, we present the pressure of neutron star matter as functions of
energy density. Comparing with nuclear matter (solid gray line), it is found that the EoS is softened once $\Lambda$-hyperon
appears (shaded area) at approximately 300~MeV~fm$^{-3}$. With larger $\alpha_{\omega\Lambda}$, the EoS becomes stiffer due to the increasing repulsive interaction
which originated from $\omega$-mesons at larger densities.

\begin{figure}
\includegraphics[width=\linewidth]{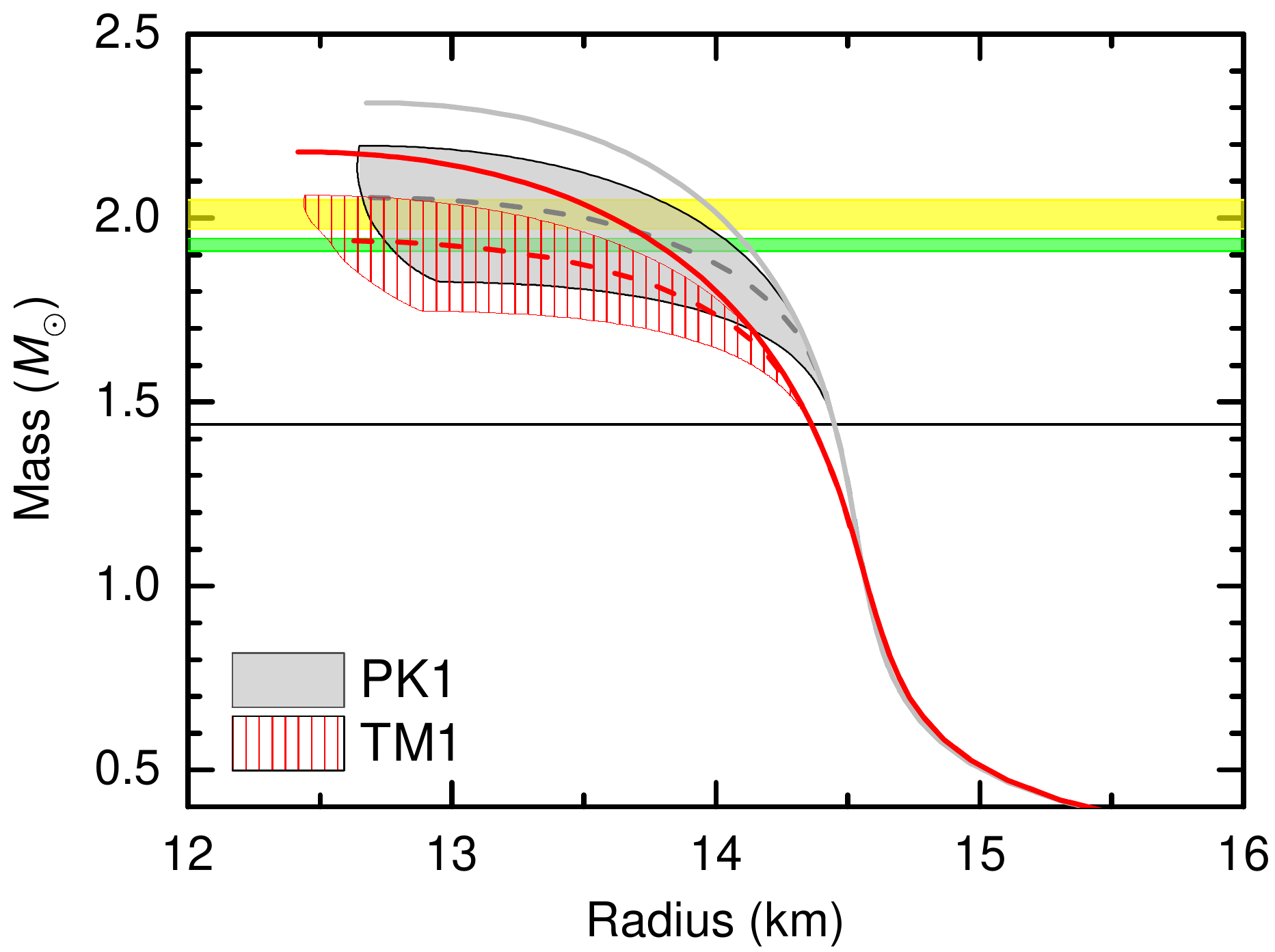}
\caption{\label{Fig:MR} (Color online) The obtained $M$-$R$ relations of neutron stars including the possible existence of $\Lambda$-hyperons.
The masses of pulsars PSR J1614-2230 ($1.928 \pm 0.017\ M_\odot$)~\cite{Demorest2010_Nature467-1081, Fonseca2016_ApJ832-167},
PSR J0348+0432 ($2.01 \pm 0.04\ M_\odot$)~\cite{Antoniadis2013_Science340-6131}, and PSR 1913+16 ($1.4398 \pm 0.0002\
M_\odot$)~\cite{Weisberg2010_ApJ722-1030} are indicated with horizonal bands.}
\end{figure}

Now the structures of neutron stars can be determined by solving the TOV equation in Eq.~(\ref{eq:TOV}) based on the EoSs displayed in
Fig.~\ref{Fig:EoS}. At the densities below half of the nuclear saturation density ($\sim 0.08\ \mathrm{fm}^{-3}$), we adopt the EoS given in
Refs.~\cite{Feynman1949_PR75-1561, Baym1971_ApJ170-299, Negele1973_NPA207-298} since uniform nuclear matter becomes unstable and a transition
to crystalized matter forming the neutron star crust takes place. In Fig.~\ref{Fig:MR}, we present the obtained $M$-$R$ relations of
neutron stars with possible existence of hyperonic matter. If we take the commonly-used value of $\alpha_{\omega\Lambda} = 0.666$ according
to the naive quark model~\cite{Dover1984_PPNP12-171}, the obtained maximum neutron star mass is lower than $1.8\ M_\odot$, in disagreement
with the observed masses of PSR J1614-2230 and PSR J0348+0432 ($\sim 2\ M_\odot$). We note, however, that only a slight adjustment of
$\alpha_{\omega\Lambda}$ is necessary to reach the observed masses of these pulsars, specifically $\alpha_{\omega\Lambda} \gtrsim 0.8$
for PK1 and $\alpha_{\omega\Lambda} \gtrsim 0.9$ for TM1. Correspondingly, the $\sigma$-$\Lambda$ couplings should be larger, i.e.,
$\alpha_{\sigma\Lambda} \gtrsim 0.73$ for PK1 and $\alpha_{\sigma\Lambda} \gtrsim 0.8$ for TM1, to obtain agreement between our predictions
and the pulsar observations.

\section{\label{sec:con}Summary}
The possible existence of $\Lambda$-hyperons in neutron stars were explored in RMF models, in which we use the effective interactions
PK1 and TM1 for nucleons while the $\Lambda$-meson couplings were constrained according to the experimental single-$\Lambda$ binding
energies of $\Lambda$-hypernuclei. A simple relation between the $\Lambda$-meson couplings ($g_{\omega\Lambda}$ and $g_{\sigma\Lambda}$)
was obtained which gives a constant value for the depth of $\Lambda$ potential $V_\Lambda = -29.786$~MeV for symmetric nuclear matter at
saturation densities. With these baryon-meson couplings, we studied the properties of neutron star matter including $\Lambda$-hyperons
and found that the onset densities for $\Lambda$ lie in the range of $0.30$-$0.36\ \mathrm{fm}^{-3}$. The EoSs were softened
at densities above which $\Lambda$-hyperons appear. However, the EoS is less softened if we adopt larger values of $\alpha_{\omega\Lambda}=
g_{\omega\Lambda}/g_{\omega N}$. By solving the TOV equation with new EoSs, we found that the maximum mass of neutron stars can reach
2 $M_\odot$ if we use $\alpha_{\omega\Lambda} \gtrsim 0.8$ for PK1 and $\alpha_{\omega\Lambda} \gtrsim 0.9$ for TM1, values just slightly
higher than those used in the naive quark model. Thus, we conclude that the values of $\Lambda$-meson couplings should be close to those
of nucleon-meson couplings so that the single-$\Lambda$ binding energies agree with measured data and the maximum mass of neutron stars
is consistent with the latest observed masses of pulsars.

Furthermore, concerning the recently observed neutron star merger gravitational wave (GW170817)~\cite{Abbott2017_PRL119-161101}
and the possible gravitational wave signal of a post-merger remnant~\cite{LIGOVirgo2017_arXiv1710.09320}, the EoSs and $\alpha_{\omega\Lambda}$
may be even further constrained. Meanwhile, heavier hyperons such as $\Xi^-$, which possibly play important roles at large densities, have not been considered.
The effects of $\Xi^-$ to the properties of neutron star matter, and accordingly to the structure of compact stars, should be considered in
our future works.

\section*{ACKNOWLEDGMENTS}
T.-T. S. and C.-J. X. express great thanks to Dr.~H.~Togashi for helpful suggestions and comments. C.-J. X. is grateful to
Prof. H.~Shen for for fruitful discussions. This work was supported by National Natural Science Foundation of China
(Grant Nos.~11525524, 11505157, 11375022, 11705163, and 11621131001), National Key Basic Research Program of
China (Grant No.~2013CB834400), the Physics Research and Development Program of Zhengzhou University (Grant No.~32410017), and the Office of Nuclear Physics in the U.S. Dept. of Energy.
The computation of this work was supported by the HPC Cluster of SKLTP/ITP-CAS and the Supercomputing Center, CNIC of CAS.


%

\end{document}